\documentclass[reprint,twocolumn,aps,prl]{revtex4} 
\usepackage{graphicx}                  
\usepackage[version=3]{mhchem}   
\usepackage{siunitx}                         

\begin{document}

\title{Charge order in \ce{SrCoO_{3-y}} investigated through resonant soft x-ray scattering}

\author{F. J. Rueckert$^{1}$, F. He$^{2}$, B. Poudel$^{3}$, B. Dabrowski$^{3}$, W. A. Hines$^{4}$, J. I. Budnick$^{4}$, and B. O. Wells$^{4}$ }
\affiliation{$^{1}$Department of Sciences, Wentworth Institute of Technology, Boston, MA 02115\\
$^{2}$Canadian Light Source, 44 Innovation Boulevard, Saskatoon, SK, Canada\\
$^{3}$Department of Physics, Northern Illinois University, DeKalb, Illinois 60115, USA\\
$^{4}$Department of Physics, University of Connecticut, Storrs, CT 06269, USA}

\date{\today}

\begin{abstract}
Resonant soft X-ray scattering was used to determine the presence of more subtle orderings not detected in standard structural analyses. By tuning to specific Co absorption edges, arrangements particular to the electronic states of those elements are enhanced. We have discovered an ordering commensurate to the lattice at the ($\frac{1}{4}~\frac{1}{4}~\frac{1}{4}$) position. Incommensurate peaks near this position were also observed. The intensity of these peaks depends on the oxygen concentration of the sample, and can be suppressed at temperatures above $\SI{320}{\kelvin}$. Regular orderings of charge density which closely match the underlying lattice may help to explain the observed propensity for \ce{SrCoO_{3-y}} (0 $\leq$ y $\leq\frac{1}{2}$) to stabilize at particular phases.
\end{abstract}

\maketitle

Charge and spin order were detected in $x=\frac{1}{8}$ doped \ce{(La$,$Nd)_{2-x}Sr_{x}CuO_{4} } and \ce{La_{2-x}Ba_{x}CuO_{4} } by neutron scattering in the 1990s \cite{Tranquada1995}. Later, resonant soft x-ray scattering confirmed the ordering involves doped holes \cite{abbamonte2005}. Recently, there has been a flurry of observations of charge density wave order in many cuprate compounds \cite{Ghiring, daSilva, Comin}, and a push to understand what this might mean for high temperature superconductivity \cite{Kivelson2014, Davis2013, Harrison2011}.  The $\frac{1}{8}$ charge order was also observed in 3-D manganites \ce{La_{1-x}Sr_{x}MnO3} \cite{KawanoPRB1996, PinsardPhysicaB1997, DabrowskiPRB1999}. An important question is whether such ordering is intimately connected to superconductivity itself or a more general feature of a doped Mott Insulator. If the latter, we might expect to find charge ordering in a range of poorly conducting, transition metal oxide doped insulators \cite{Dagotto2005, popovic2002}. Here we report on a system with similar ordering,  \ce{SrCoO_{3-y}}. 

\ce{SrCoO_{3-y}} and its related compounds \ce{Sr_{1-x}La_{x}CoO_{3}} form a materials system with a phase diagram in some ways similar to the cuprates. In both cases, cation doping leads to short length scale inhomogeneity while anion doping leads to large length scale phase separation. For the cuprates, this behavior is well documented for \ce{La2CuO4} doped with Sr or excess oxygen \cite{saviciPRB2002, JorgensenPRB1988, Mohottala2006}. In the cobaltates, \ce{LaCoO3} is a correlated insulator with \ce{Co^3+} but has a spin state transition with no moment at low temperatures \cite{Magnuson2004}. As Sr is introduced onto La sites, the low spin state is quickly suppressed and a nanoscale magnetically phase separated region emerges consisting of ferromagnetic islands in a paramagnetic background. At a percolation threshold of \ce{La_{0.67}Sr_{0.33}CoO3} long range ferromagnetism and a conducting state emerge \cite{HeEuro2009}. The Curie temperature then continuously increases for greater Sr concentrations approaching \ce{SrCoO3} \cite{HePRB2007}. 

\ce{SrCoO_{2.5}} is a Mott insulator with nominally \ce{Co^3+} and a Neel temperature of $\SI{525}{\kelvin}$ \cite{Takeda1972}. It forms in the Brown-Millerite structure, an oxygen vacancy ordered variant of the perovskite lattice. Oxygen can be added to the lattice which increases the Co valence. At low oxygen concentrations, the material will phase separate between the Mott insulator phase of \ce{SrCoO_{2.5}} and ferromagnetic \ce{SrCoO_{2.75}}. Further oxygen can be added, though to reach stoichiometric \ce{SrCoO3} strong oxidation methods such as electrochemistry are needed. At these high oxygen concentrations, the oxygen itself appears to be continuously soluble, but samples spontaneously break into separate magnetic regions that match those at the special concentrations of \ce{SrCoO_{2.75}}, \ce{SrCoO_{2.88}}, and \ce{SrCoO3} with T$_\textrm{C}$ of $\SI{160}{\kelvin}$, $\SI{220}{\kelvin}$, and $\SI{280}{\kelvin}$, respectively \cite{Xie2011}. Further work using muon spin rotation and neutron powder diffraction confirmed that the mixed phase samples consist of spatially separated, distinct magnetic phases with an intermediate length scale of several tens of nanometers \cite{ZZhuPRB}. 

In both cuprate and cobaltate systems, as the correlated insulator is chemically doped on either cation or anion doping axis, the antiferromagnetic insulating state is quickly killed, and ferromagnetic phases appear. The discovery of the almost ubiquitous presence of charge order in the cuprates was made possible through the use of resonant soft x-ray scattering. In this manuscript we use the same technique to reveal a previously unknown ordering in oxygen deficient samples of \ce{SrCoO_{3-y}}. The resonant scattering is very strong, and particular orderings appear to define particularly stable charge states for these compounds. 

\begin{figure}
\includegraphics[width=3in]{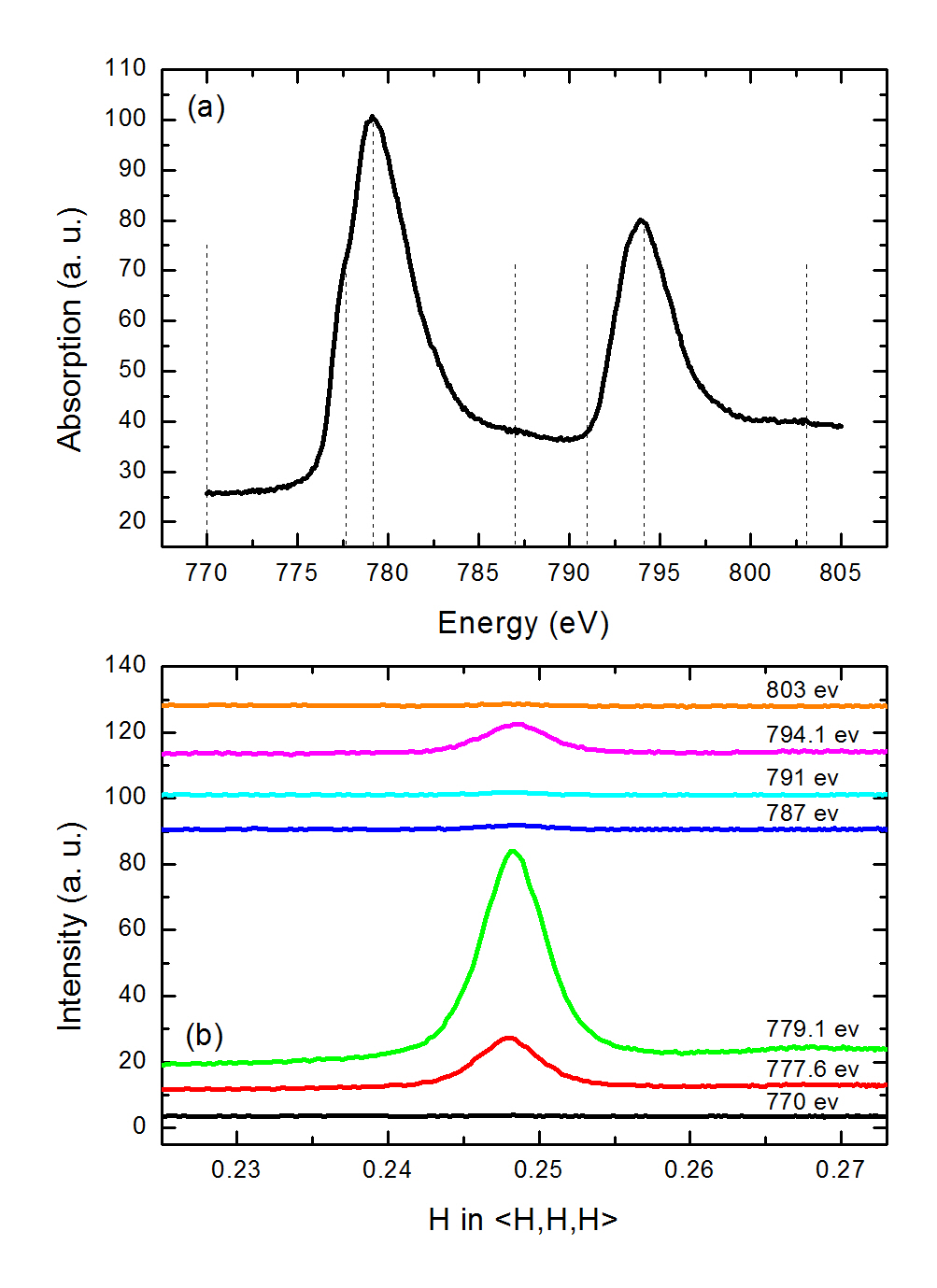}  
\caption{(Color online) (a) Absorption profile of $y=\frac{1}{8}$ film taken at $T~=~\SI{250}{\kelvin}$ (T$_\textrm{C}$ = $\SI{220}{\kelvin}$). (b) Wide angle XRD measurement for each of the energies noted in panel (a). Peaks are evident only at the L$_{3}$ and L$_2$ edges.}
\label{fig1}
\end{figure}

Figure \ref{fig1} presents an example of the basic finding of this work. We began a search for resonant superlattice peaks using powder samples synthesized to a stable oxygen concentration of \ce{SrCoO_{2.88}}.  We identified a very weak peak with a spacing near  ($\frac{1}{4}~\frac{1}{4}~\frac{1}{4}$) with reference to the perovskite unit cell. To create a sample more suitable for scattering studies we proceeded to grow epitaxial films on \ce{SrTiO3} substrates, with those used here oriented in the $\langle 1~1~1 \rangle$ direction.  As we have previously reported \cite{FJR_APL}, as grown and annealed films have an oxygen concentration of \ce{SrCoO_{2.75}}, but are easily electrochemically oxidized to \ce{SrCoO_{2.88}}, a phase that is metastable over weeks. 

Figure \ref{fig1}a shows the absorption spectrum of the \ce{SrCoO_{2.88}} film over the Co L$_{2}$ and L$_{3}$ edges with data collected in the total fluorescence mode. The absorption edges are fairly broad, with the peak absorption near $\SI{779}{\electronvolt}$ and $\SI{794}{\electronvolt}$ for the L$_{3}$ and L$_{2}$ edges, respectively. The L$_{3}$ edge has a leading edge shoulder near $\SI{777.6}{\electronvolt}$ with perhaps a similar feature near $\SI{793}{\electronvolt}$ for the L$_{2}$ edge. Figure \ref{fig1}b reproduces a set of longitudinal x-ray diffraction (XRD) scans taken with varying photon energies matching the dashed lines in part a. For energies away from the absorption edge, such as the bottom scan taken at $\SI{770}{\electronvolt}$, no peak is discernible. However, a peak near ($\frac{1}{4}~\frac{1}{4}~\frac{1}{4}$) appears at a photon energy corresponding to the leading shoulder of the L$_{3}$ edge that becomes very intense at the peak energy of the L$_{3}$ absorption. At higher energies the peak again disappears, then resonates much more weakly over the L$_{2}$ absorption edge. 

Unfortunately, the limited momentum of photons at these low energies does not allow a full characterization of the structures responsible for this resonant diffraction. We can determine that this peak represents an ordering associated with valence states of the Co ions indicating a large ordered super-cell. Such an ordering has not been previously reported in the Co-perovskite systems. The strength of the resonance appears much larger than that reported in the cuprates \cite{abbamonte2005}, and likely as large as that seen in the nickelates \cite{SchusslerPRL2005}. 

\begin{figure}
\includegraphics[width=3in]{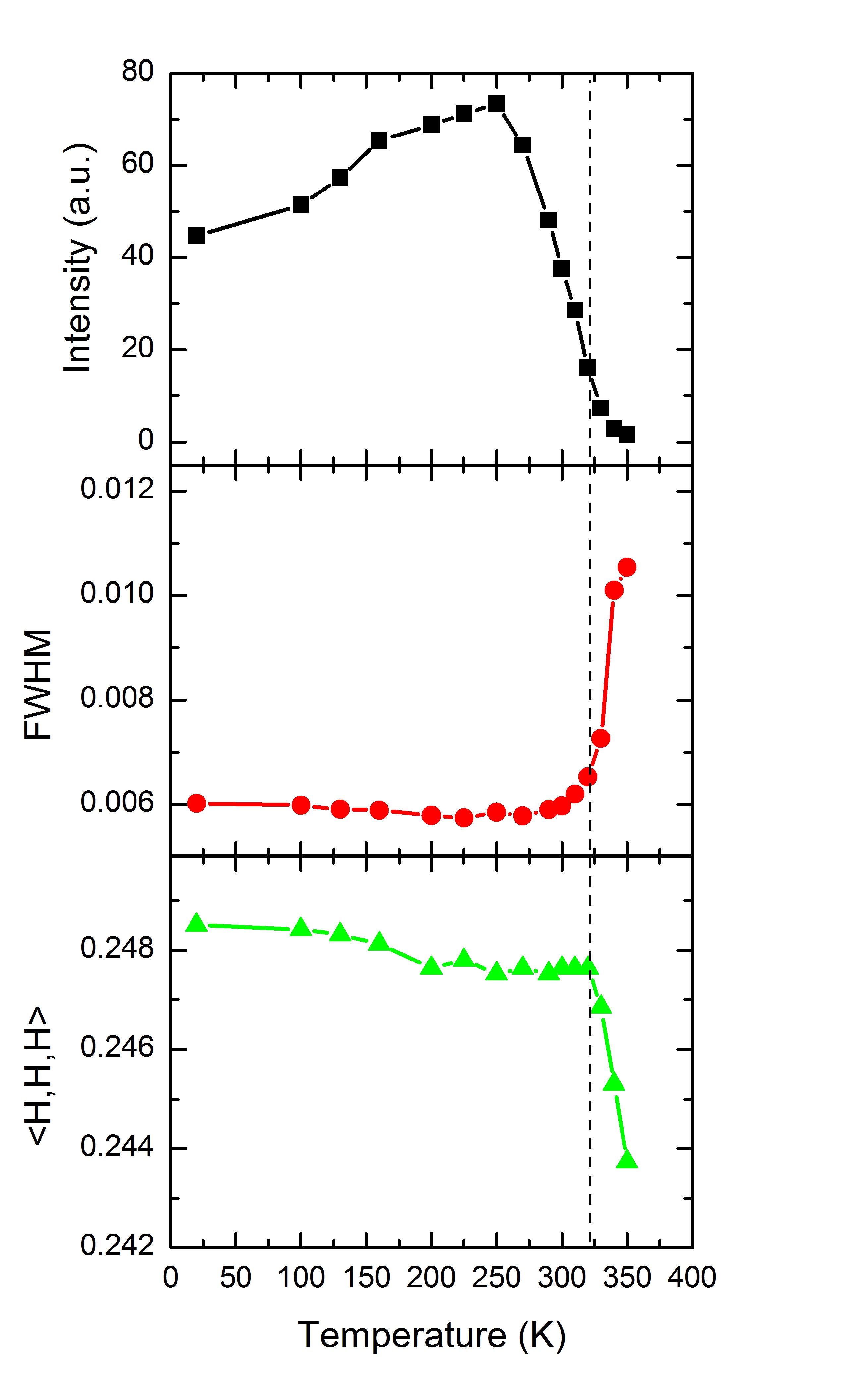}  
\caption{(Color online) Details of the commensurate ($\frac{1}{4}~\frac{1}{4}~\frac{1}{4}$) peak for the 300 nm thick $y=\frac{1}{8}$ film as a function of temperature. Intensity is shown in arbitrary units. Full-width half-max and position are measured in reciprocal lattice units along the $\langle 1,1,1 \rangle$ direction. T$=\SI{320}{\kelvin}$ (dashed line) represents an ordering transition above which the ordering peak is suppressed.}
\label{fig2}
\end{figure}

 Figure \ref{fig2} shows the temperature dependence of the ($\frac{1}{4}~\frac{1}{4}~\frac{1}{4}$) peak taken upon warming the \ce{SrCoO_{2.88}} sample from $\SI{20}{\kelvin}$ to $\SI{360}{\kelvin}$.  The top panel gives the intensity as a function of temperature. While the intensity increases from $\SI{20}{\kelvin}$ to $\SI{250}{\kelvin}$, it is unclear if the gain is due to dynamics, or an actual diminution of the order at low temperatures. At higher temperatures the peak disappears as expected for an ordering transition. The middle panel gives the peak width as a function of temperature. This is constant below the ordering temperature but diverges above, as is consistent with a second order phase transition. The actual width at low temperature, approximately 0.006 reciprocal lattice units along $\langle 1~1~1 \rangle$, represents a lower bound on the domain size of the related ordering of approximately 150 unit cells, or $\SI{1000}{\angstrom}$. The lower panel is the peak position as a function of temperature. It varies somewhat strongly in the region of critical scattering above the transition, but locks in at a nearly commensurate position at the transition.

\begin{figure}
\includegraphics[width=3in]{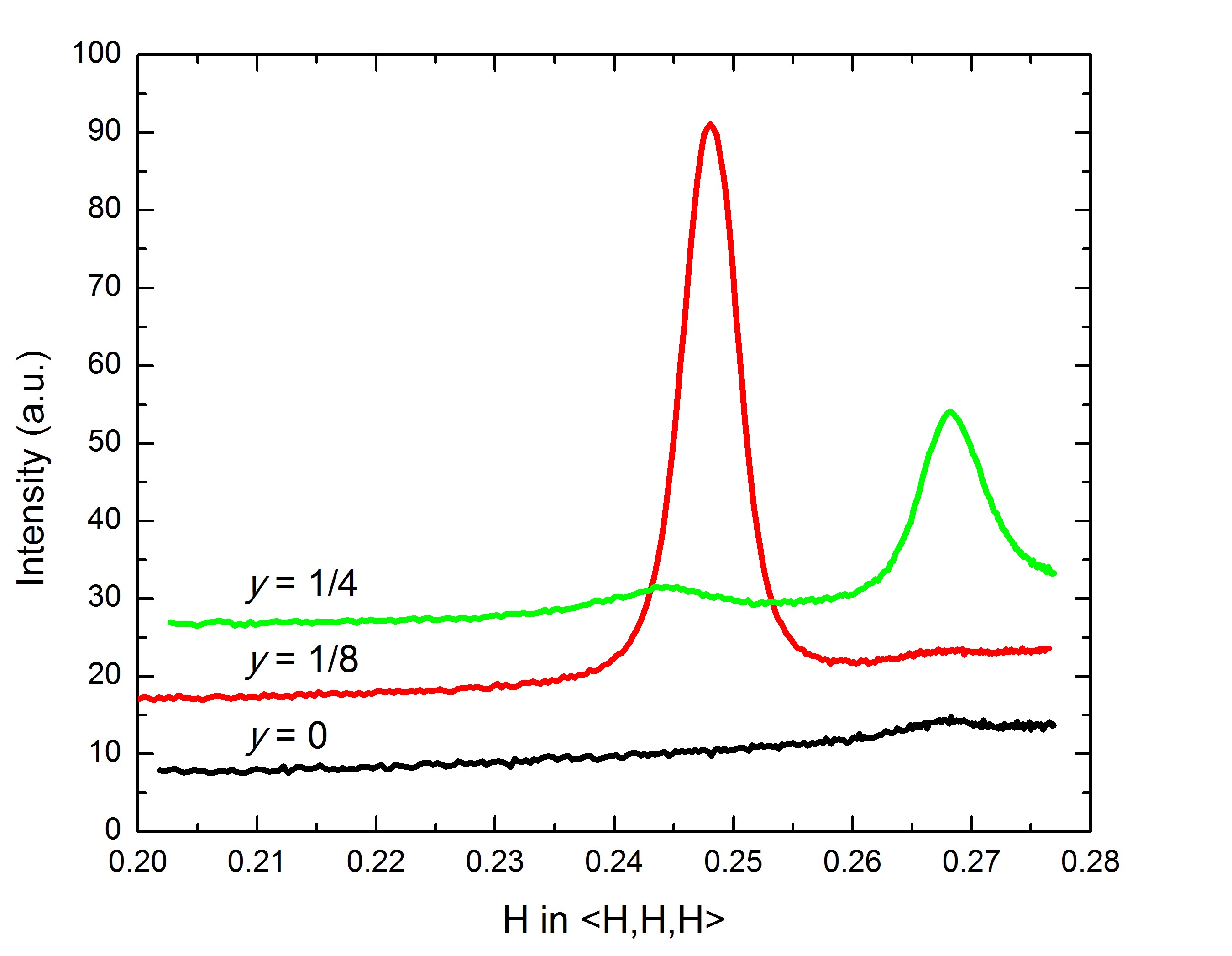}  
\caption{(Color online) Wide angle XRD scans showing the positions of the ordering peaks for each oxygen concentration. The scans were taken below the onset of charge ordering at $\SI{320}{\kelvin}$. Oxygen concentrations $y=\frac{1}{4}, \frac{1}{8},$ and $0$, were taken at temperatures of $200, 250,$ and $\SI{20}{\kelvin}$, respectively. Lines are displaced for clarity.}
\label{fig3}
\end{figure}

To produce a careful study of the behavior of this peak as a function of doping, we investigated a single film sample at various stages of oxidation. We determine the oxygen content for the film by comparing lattice constants and magnetic transitions to those of the bulk samples. The data set of Fig.\ref{fig3} are x-ray diffraction (XRD) scans taken at 779 eV from a 300 nm thick film initially in the as-grown \ce{SrCoO_{2.75}} state, then in a fully electrochemically oxidized \ce{SrCoO3} state. Finally, after some time in vacuum, the film reverts to the \ce{SrCoO_{2.88}} state we had previously measured. The fully oxidized sample showed only a very small peak on resonance, with no temperature dependence, which we attribute to a small portion of the sample not oxidized, similar to what has been observed in super-oxygenated cuprates \cite{Wells96}. After losing oxygen over the course of several hours in vacuum, a very strong peak developed at the ($\frac{1}{4}~\frac{1}{4}~\frac{1}{4}$) position, matching that seen previously for \ce{SrCoO_{2.88}} samples. The sample with the lowest concentration of oxygen, the as-grown \ce{SrCoO_{2.75}} showed two peaks in this region of reciprocal space, both along the $\langle 1~1~1 \rangle$ direction. The weaker of the two peaks occurs fairly close to a commensurate position at ($0.244~0.244~0.244$) while the more intense peak is near ($0.27~0.27~0.27$). 
 
Since the superlattice reflections disappear for the fully oxygenated \ce{SrCoO3} sample, we conclude that the ordering is associated with variations caused by the missing oxygen. One consideration is the possibility of oxygen vacancy ordering. However, since the resonance is associated with the Co absorption edge, the direct cause of the scattering must be due to variations in the Co ions. For \ce{SrCoO_{2.88}} we expect one oxygen vacancy for every eight unit cells, which we might associated with a mixture of \ce{Co^3+} and \ce{Co^4+} at a ratio of $\frac{1}{4}:\frac{3}{4}$.
  
The \ce{SrCoO_{2.75}} sample does not seem to fit into the same simple picture. Rather than the single commensurate peak for the \ce{SrCoO_{2.88}} sample, two incommensurate peaks are observed. This appears to reflect a much more complicated arrangement of Co valence states that is not possible to directly identify.

Taking all of the data together, there appears to be a somewhat rounded transition with an ordering temperature near $\SI{320}{\kelvin}$. While we have not performed a full hysteretic analysis of this transition, we have found that upon cooling the sample from $\SI{360}{\kelvin}$ to $\SI{300}{\kelvin}$, the peak reforms with the same line shape and intensity, representative of an ordering transition.  This ordering is distinct from the ferromagnetic order which sets in near $\SI{220}{\kelvin}$. Thus we have identified a new transition not previously found, associated with the ordering of \ce{Co^3+} and \ce{Co^4+} ions.

\begin{figure}
\includegraphics[width=3in]{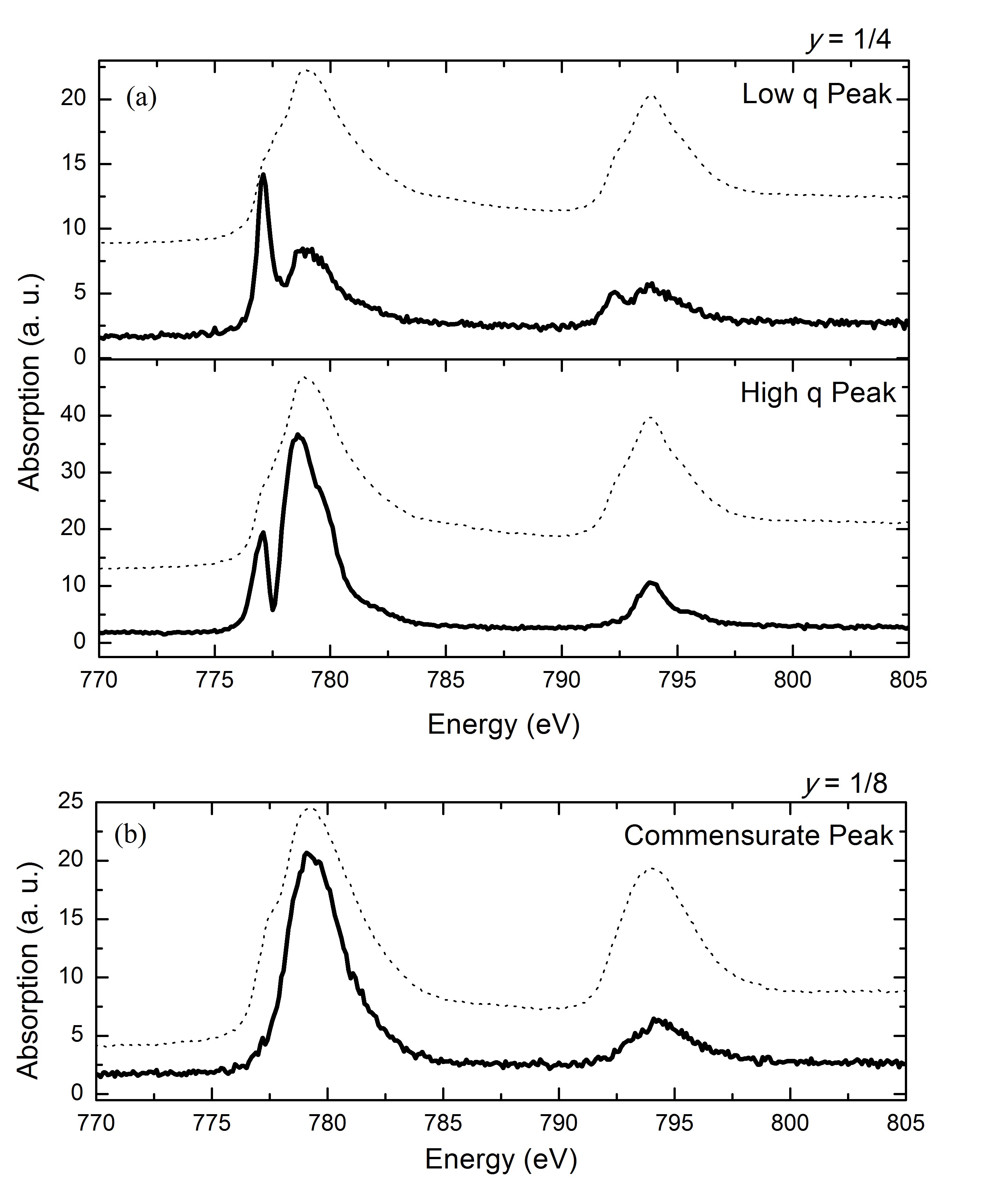}  
\caption{Absorption profile of the ordering peaks, taken at T = $\SI{250}{\kelvin}$. (a) Measurement of the incommensurate peaks for $y=\frac{1}{4} $ at low and high q positions. (b) Profile for the $y=\frac{1}{8}$ phase at the commensurate peak. The dotted lines show the background absorption taken off the peak and scaled for comparison.}
\label{fig4}
\end{figure}

In addition to traditional q-scans of the diffraction peaks, the energy dependence of the diffraction peak intensity was measured. Figure \ref{fig4} shows the energy dependence of the three peaks that appear in Fig.\ref{fig3}. The experimental procedure scans the incident energy and compares the intensity with the detector fixed at different positions in q. The difference between scans fixed on and well off the peak gives the energy profile of the scattering peak.

The absorption profiles are shown in Fig.\ref{fig4}. The two scans in panel (a) give the energy dependence of the two peaks of the as grown \ce{SrCoO_{2.75}} sample from Fig.\ref{fig3}. The scans are labelled by position as low q (also lower intensity) and high q (the more intense peak). Panel (b) shows the energy profile of the single peak from the \ce{SrCoO_{2.88}} sample from Fig.\ref{fig3}. The background absorption is shown in each figure as a dotted line for comparison.

A notable difference occurs near $\SI{778}{\electronvolt}$. For the $y=\frac{1}{4}$ phase, the absorption maximum occurs at a prepeak feature for the low q position, while the maximum for the high q peak occurs at the L$_{3}$ edge. A similar phenomenon occurs at the L$_{2}$ edge. The absorption behaviors of the two peaks at this phase are the inverse of each other. For the $y=\frac{1}{8}$ phase, no such feature exists in the absorption. The lack of correspondence between the absorption profiles for the near commensurate positions of each phase may indicate a fundamental difference between the types of ordering in each instance and implies different valence states are involved.

Our data provide a reasonable picture for finding a possible driving force for electronic phase separation in \ce{SrCoO_{3-y}} previously observed \cite{Xie2011, FJR_APL}. A unique ordering pattern is associated with each stable phase. The picture is simplest for materials with oxygen content between \ce{SrCoO_{2.88}} and \ce{SrCoO3}. By electron counting, \ce{SrCoO3} should consist of uniformly \ce{Co^4+} ions. Thus there are no species to provide a superstructure order and we detect no related scattering peaks. For \ce{SrCoO_{2.88}} we expect a mixed Co valence, with $\frac{1}{4}$th of the Co ions reduced to \ce{Co^3+}. Allowing such charge ordering to be commensurate with the lattice would logically provide a low energy configuration, and our result gives evidence for just such an ordering.

The energy profile of the scattering peak in \ce{SrCoO_{2.88}} is generally close to that of the overall x-ray absorption spectrum, though it lacks the shoulder at lower energy. The shoulder itself appears related to \ce{Co^4+}, as it is more prominent in \ce{SrCoO3} than the samples with less oxygen. In addition the low energy shoulder is shown to be associated with \ce{Co^4+} in calculations in a previous x-ray absorption spectroscopy (XAS) study of \ce{NaCoO_{x}} where Co can take on 3+ and 4+ oxidation states \cite{Hu_76}. However, the scattering peak resonates primarily over the main part of the absorption edge. This is much like the case in \ce{La_{1.875}Ba_{0.125}CuO4} or \ce{La_{1.475}Nd_{0.4}Sr_{0.125}CuO4} \cite{abbamonte2005,AchkarPRL2013}. Here, recent work has attributed the source of resonant contrast as a local shift in energy of the absorption profile associated with sites of differing valence \cite{AchkarPRL2013}.

The situation for \ce{SrCoO_{2.75}} is not so simple. There is a specific ordering of Co ions associated with the stable \ce{SrCoO_{2.75}} phase. The ordering is more complex than in \ce{SrCoO_{2.88}} as reflected by the two scattering peaks, neither of which are commensurate. The energy profiles of the peaks, shown in Fig.\ref{fig4}a, have rich structures and are complementary. The low q peak resonates most strongly in the pre-peak shoulder region of the XAS while the high q peak shows the inverse energy dependence: a stronger resonance on the main part of the absorption. At this point we cannot derive a definitive model for the ordering in \ce{SrCoO_{2.75}}, which will be a topic for further work. However, it is clear that the stable phase is associated with a particular set of ordering peaks. 
 
There are clear parallels between this work and many recent reports of charge ordering in cuprate superconductors. Firstly, one should note that the samples themselves can be similarly described as strongly correlated, off-stoichiometry compounds. The cuprates are well known to have Mott insulating parent compounds which can be made superconducting by altering the stoichiometry such that holes are introduced into the Cu-O layers. Samples doped with about 0.16 holes per Cu site have maximum T$_\textrm{c}$ for their particular family whereas those with a doping of 0.125 holes per Cu site have suppressed superconductivity and in many cases charge ordering as revealed by soft x-ray resonant diffraction \cite{Ghiring, daSilva, Comin}. In this case, \ce{SrCoO_{2.88}} can be considered either electron doped versus fully stoichiometric \ce{SrCoO3} or, in stronger parallel to the cuprates, as heavily hole doped from the Mott insulator \ce{SrCoO_{2.5}} \cite{Takeda1972}. For most cuprates, the charge-ordering peaks are extremely hard to measure when not employing the strong resonance condition, and the resonance can be seen at both the Cu L edge and the O K edge. Similarly, in \ce{SrCoO_{2.88}} the resonance is not easy to detect in standard diffraction but resonates strongly at the Co L edge. However, for \ce{SrCoO_{2.88}}, the q-value of the resonant peak is such that the scattering triangle cannot be closed at the O K edge.

Thus this work indicates that charge ordering may be a broad feature of doped Mott insulators. In fact, charge ordering itself may provide a significant electronic condensation energy leading to stable charge densities and even electronic phase separation. In this work, the stable phases associated with \ce{SrCoO_{2.88}} and \ce{SrCoO_{2.75}} have specific, different charge ordering arrangements. This is similar to the $\frac{1}{8}$th doped phase in the cuprates which is quite stable and now almost universally associated with charge ordering \cite{Tranquada1995,abbamonte2005,Ghiring,daSilva,Comin}. Long length scale phase separation may require other favorable conditions. This is seen in \ce{SrCoO_{3-y}} and \ce{La2CuO_{4+y}} \cite{saviciPRB2002,Mohottala2006} which have both charge/spin order and mobile oxygen defect-dopants \cite{PaulusJACS2008,PaulsJSSE2011}.

Research described in this paper was performed at the Resonant Elastic and Inelastic X-ray Scattering \mbox{(REIXS)} Beamline of the Canadian Light Source, which is supported by the Canada Foundation for Innovation, Natural Sciences and Engineering Research Council of Canada, the University of Saskatchewan, the Government of Saskatchewan, Western Economic Diversification Canada, the National Research Council Canada, and the Canadian Institutes of Health Research. Sample synthesis at the University of Connecticut was supported by the NSF through Grant No. DMR-0907197, data collection and analysis by DOE-BES Contract No. DEFG02-00ER45801. Research at NIU was supported by the Great Journeys Assistantship, Northern Illinois University.

\bibliographystyle{iopart-num}
\bibliography{References}

\end{document}